\documentstyle[12pt,amssymb,aps,epsf]{revtex}

\def\PRL#1#2#3{{\sl Phys. Rev. Lett.} {\bf#1} (#2) #3}
\def\PR#1#2#3{{\sl Phys. Rev.} {\bf#1} (#2) #3}

\def\NPB#1#2#3{{\sl Nucl. Phys.} {\bf B#1} (#2) #3}

\def\PRB#1#2#3{{\sl Phys. Rev.} {\bf B#1} (#2) #3}

\def\PLA#1#2#3{{\sl Phys. Lett.} {\bf #1A} (#2) #3}
\def\JMP#1#2#3{{\sl J. Math. Phys.} {\bf #1} (#2) #3}

\def\PTP#1#2#3{{\sl Prog. Theor. Phys.} {\bf #1} (#2) #3}

\def\RMP#1#2#3{{\sl Rev. Mod. Phys.} {\bf #1} (#2) #3}

\def\IJMPB#1#2#3{{\sl Int. J. Mod. Phys.} {\bf B#1} (#2) #3}

\def\AdP#1#2#3{{\sl Advances in Phys.} {\bf #1} (#2) #3}

\def\TMP#1#2#3{{\sl Theor. Mat. Phys.} {\bf #1} (#2) #3}

\def\JPA#1#2#3{{\sl J. Physics} {\bf A#1} (#2) #3}
\def\JPC#1#2#3{{\sl J. Physics} {\bf C#1} (#2) #3}

\def\JSM#1#2#3{{\sl J. Soviet Math.} {\bf #1} (#2) #3}

\begin{document}
\def\mb{\bar{\mu}}
\def\da{\downarrow}
\def\up{\uparrow}
\def\be{\begin{equation}}
\def\ee{\end{equation}}
\def\bea{\begin{eqnarray}}
\def\eea{\end{eqnarray}}
\def\eps{\varepsilon}
\def\g{\gamma}
\def\b{\beta}
\def\d{\delta}
\def\a{\alpha}
\def\e{\varepsilon}
\def\l{\lambda}
\def\lp{\lambda^\prime}
\def\o{\omega}
\def\s{\sigma}
\def\La{\Lambda}
\def\nn{\nonumber\\}
\def\gt{\widetilde{G}}
\def\G{\Gamma}
\def\w{\langle W|}
\def\tA{\tilde A}
\def\tB{\tilde B}
\def\tr{\widetilde{\rho}}
\def\vr{\varrho}
\def\v{|V\rangle}
\def\wt{\langle {\tilde W}|}
\def\vt{|{\tilde V}\rangle}
\def\r#1{(\ref{#1})}
\def\sm{{\bar Q}}
\def\cq{{\cal Q}}
\def\t#1{\langle\tau_{#1}\rangle}
\def\2t#1#2{\langle\tau_{#1}\tau_{#2}\rangle}
\def\up{\uparrow}
\def\1l{\lambda^{(1)}}
\def\lp{\Lambda_p}
\def\lh{\Lambda_h}
\def\ph{\lp,\lh}
\begin{titlepage}
\begin{flushleft}
May 1996
\end{flushleft}
\thispagestyle{empty}
\begin{center}
\vspace*{1cm}
{\Large\sc The Supersymmetric $t$-$J$ Model with a Boundary}\\[10mm]
{ {\sc Fabian H.L. E\char'31ler}\footnote{%
e-mail: {\tt fab@thphys.ox.ac.uk}}}
\\[9mm]

\begin{center}
{\small\sl Department of Physics, Theoretical Physics\\
     Oxford University\\
     1 Keble Road, Oxford OX1 3NP, Great Britain}\\[8pt]
\end{center}
\vspace*{1cm}
ABSTRACT
\end{center}
\begin{quote}
\baselineskip=14pt
An open supersymmetric $t$-$J$ chain with boundary fields is studied
by means of the Bethe Ansatz. 
Ground state properties for the case of an almost half-filled band and
a bulk magnetic field are determined. Boundary susceptibilities are
calculated as functions of the boundary fields. The effects of the
boundary on excitations are investigated by constructing the exact
boundary S-matrix. From the analytic structure of the boundary
S-matrices one deduces that boundary bound states are formed for
sufficiently strong boundary fields.
\end{quote}

\vspace*{.5cm}


\hfill
\end{titlepage}
\baselineskip=14pt
\section{\sc Introduction}
Recently there has been renewed interest in one-dimensional impurity
problems and the related problem of one-dimensional Luttinger liquids
with boundaries \cite{tsv,tsv0,bares0,saleur,henrik,fls,kf,aff}.
The main focus of these investigations has been the effects of
Kondo-like impurities and effects due to potential scattering in
Luttinger liquids. These impurity problems are closely related to open
1-d systems with boundary fields. Some of these systems are integrable
and can be solved exactly by Bethe Ansatz
\cite{che,xxz2,skl,mn1,xxz1,schulz}. In particular, in \cite{tsv} an
anisotropic Heisenberg chain with open boundary conditions was studied.
It is the purpose of the present work extend the investigation of
\cite{tsv} to the case of the $t$-$J$ model, which is a Luttinger
liquid with both spin and charge degrees of freedom. 

In \cite{gr} a trigonometric generalization of the supersymmetric
$t$-$J$ model with open boundaries was constructed by means of the
Quantum-Inverse Scattering Method (see {\sl e.g.} \cite{vladb}). This
generalized the previous work by F\"orster and Karowski on the
quantum-group invariant case \cite{fk2}. Here we study this model at
the rational point, for which it reduces to the supersymmetric $t$-$J$
model with open boundaries and boundary fields. The reason for this
restriction is that the trigonometric model in general leads to a
non-hermitian bulk hamiltonian. The only exception is the hyperbolic
regime, but there the spin excitations are gapped and thus irrelevant
for the low-energy physics of the model. 
The hamiltonian we consider in the grand canonical ensemble is given
by
\bea
H&=& -{\cal P}\left(\sum_{j=1}^{L-1}\sum_\sigma c^\dagger_{j,\sigma}
c_{j+1,\sigma}+c^\dagger_{j+1,\sigma}c_{j,\sigma}\right){\cal P}\nn
&&+2\sum_{j=1}^{L-1}{\vec S_j}\cdot{\vec S_{j+1}}
-\frac{n_jn_{j+1}}{4} +\sum_{j=1}^{L-1}n_j+n_{j+1} - H S^z - \mu \hat{N} 
+H_{\a\b}\ ,
\label{hamil}
\eea
where ${\cal P}$ projects out double occupancies, $\vec{S_j}$ are spin
operators at site $j$,
$n_j=c^\dagger_{j,\up}c_{j,\up}+c^\dagger_{j,\da}c_{j,\da}$, and the
four possible choices of boundary hamiltonians $H_{\a\b}$ compatible
with integrability and conservation of total spin in z-direction  and
particle number are given by  
\bea
H_{aa}&=&h_1^\prime n_1 + h_L^\prime n_L\ ,\qquad\qquad\quad
H_{ab}=h_1 (S^z_1-\frac{n_1^h}{2}) + h_L^\prime n_L\nn
H_{ba}&=&h_1^\prime n_1 + h_L (S^z_L-\frac{n_L^h}{2})\ ,\qquad
H_{bb}=h_1 (S^z_1-\frac{n_1^h}{2}) + h_L (S^z_L-\frac{n_L^h}{2})\ .
\eea
Here $n_j^h=1-n_{j,\up}-n_{j,\da}$ is the number operator for holes
(unoccupied sites) at site $j$. To simplify the computations we
constrain ourselves to the regions $h\in (0,2)$, $h^\prime\in
(0,1)$. It is straightforward to extend the analysis below to other
ranges of the fields. For later convenience we define the quantities 
\be
S_1 = \cases{2-\frac{2}{h_1^\prime} & {\rm for} aa, ba\cr
1-\frac{2}{h_1} & {\rm for} ab, bb\cr}\ ,\quad
S_L = \cases{2-\frac{2}{h_L^\prime} & {\rm for} aa, ab\cr
1-\frac{2}{h_L} & {\rm for} ba, bb\cr}.
\label{s}
\ee
In what follows we always assume that $S_1$ and $S_L$ are noninteger
numbers. We note that for zero boundary fields \r{hamil} exhibits a
global $sl(1|2)$ symmetry \cite{fk2}.
%% : $H$ commutes with the generators
%%\bea
%%S^\dagger &=& \sum_{j=1}^{L-1} c^\dagger_{j,\up} c_{j,\da}\ , \quad 
%%S = \sum_{j=1}^{L-1} c^\dagger_{j,\da} c_{j,\up}\ , \quad 
%%S^z = {1\over 2}\sum_{j=1}^{L-1} (n_{j,\up}-n_{j,\da})\ ,\quad
%%T = \sum_{j=1}^{L-1} (1-\frac{1}{2}n_j)\ ,\nn
%%Q_{\sigma} &=& \sum_{j=1}^{L-1}(1-n_{j,-\sigma}) c_{j,\sigma}\ ,\quad
%%Q^\dagger_{\sigma} =
%%\sum_{j=1}^{L-1}(1-n_{j,-\sigma})c^\dagger_{j,\sigma}\ . 
%%\label{sl12}
%%\eea
In the present work we perform a detailed study of the boundary
effects in the model defined by \r{hamil}, paying particular attention
to the influence of the boundary fields. After some technical
preliminaries we turn to an analysis of the ground state
properties. We find that the zero-temperature susceptibilities exhibit
some interesting singularities, which we argue to be related to the
formation of bound states near the boundaries. We then study the
interaction of elementary excitations with the boundaries by computing
the exact boundary S-matrices. We find that boundary bound states can
be formed for sufficiently strong boundary fields. 
We concentrate on the case of band fillings close to one
(corresponding to the particularly interesting case of the lightly
doped Mott-Hubbard insulator) for which it is possible to obtain
explicit analytical results. However it is straightforward to extend
the present analysis to arbitrary band-fillings by solving the
integral equations \r{dresseden} numerically and then numerically
integrating \r{E}.

Taking the rational limit of the Bethe Ansatz equations derived in
\cite{gr} we obtain
\bea
\eta_{\a\b}(\l_k) \left(e_1(\l_k)\right)^{2L}&=&\prod_{j\neq k}^{N_h+N_\da}
e_2(\l_k-\l_j) e_2(\l_k+\l_j) \prod_{l=1}^{N_h} e_{-1}(\l_k-\1l_l)
e_{-1}(\l_k+\1l_l)\nn
1&=&\zeta_{\a\b}(\1l_l)\prod_{j=1}^{N_h+N_\da}e_{1}(\1l_l-\l_j)
e_{1}(\1l_l+\l_j)\ ,
\label{bae}
\eea
where $e_n(x)=\frac{x+\frac{in}{2}}{x-\frac{in}{2}}$ and
$\a,\b=a,b$. The boundary terms are given by
\bea
\eta_{aa}(\l)&=&1\ ,\quad\eta_{ab}(\l)=-e_{-S_1}(\l)\ ,\quad
\eta_{ba}(\l)=-e_{-S_L}(\l)\
,\quad\eta_{bb}(\l)=\eta_{ab}(\l)\eta_{ba}(\l)\nn 
\zeta_{bb}(\l)&=&1\ ,\quad\zeta_{ab}(\l)=-e_{-S_L}(\l)\ ,\quad
\zeta_{ba}(\l)=-e_{-S_1}(\l)\
,\quad\zeta_{aa}(\l)=\zeta_{ab}(\l)\zeta_{ba}(\l)\ .
\eea
The restrictions imposed on $h$ and $h^\prime$ are chosen such that in
all these expressions the label $x$ on $e_x(\l)$ is positive with range
$(0,\infty)$. The energy of a state corresponding to a solution of
\r{bae} is (up to an overall constant, which we drop)
\be
E=E_{ij}-\sum_{j=1}^{N_h+N_\da}\frac{1}{\frac{1}{4}+\l_j^2}
+H(N_\da+N_h)+(\mu-\frac{H}{2})N_h-(\mu+\frac{H}{2})L\ , 
\label{bareE}
\ee
where $E_{aa}=h_1^\prime+h_L^\prime$, $E_{bb}=\frac{h_1+h_L}{2}$ and
so on. The reference state used to derive \r{bae} is the one with
up-spin electrons at each site of the lattice. This leads to the
constraint in \r{bae} that the number $N_\da$ of down-spins must be
smaller than or equal to the number of up-spins $N_\up$. Solutions of
\r{bae} violating this constraint can lead to vanishing wave-functions
and must be ignored. Eigenstates of \r{hamil} with $N_\da>N_\up$ must
be constructed by switching the reference state to the state with
down-spin electrons at all sites. This leads to the same Bethe
equations \r{bae} with $N_\da\leftrightarrow N_\up$ and different
values for the quantities $S_j$
\be
S_1 = \cases{2-\frac{2}{h_1^\prime} & {\rm for} aa, ba\cr
1+\frac{2}{h_1} & {\rm for} ab, bb\cr}\ ,\quad
S_L = \cases{2-\frac{2}{h_L^\prime} & {\rm for} aa, ab\cr
1+\frac{2}{h_L} & {\rm for} ba, bb\cr}.
\ee
Below we will mainly deal with situations for which $N_\da\leq
N_\up$. However, when considering excitations over the
antiferromagnetic ground states we will also consider the case
$N_\da\geq N_\up$, for which we will make use of the procedure
outlined above.

In order to simplify \r{bae} we make use of the `string-hypothesis'
\footnote{As far as the present work is concerned we do not need to
explicitly consider complex solutions of the Bethe equations and all
of our results are really independent of the string hypothesis.},
which states that for $L\rightarrow\infty$ all solutions are composed
of real $\1l_\g$'s whereas the $\l$'s are distributed in the complex
plane according to the description
\be
\l^{n,j}_\a = \l^{n}_\a + i\left(\frac{n+1}{2}-j\right)\ , j=1\ldots n
\label{strings}
\ee
where $\a=1\ldots M_n$ labels different `strings' of length $n$. This
string hypothesis is naturally identical to the one for the model with
periodic boundary conditions. The imaginary parts of the $\l$'s can
now be eliminated from \r{bae} via \r{strings}. Taking the logarithm
of the resulting equations (for $M_n$ strings \r{strings} of length
$n$ and $N_h$ $\l^{(1)}$'s (note that $\sum_{n=1}^\infty
nM_n= N_\da+N_h$) we arrive at
\bea
\frac{2\pi}{L} I^n_\a &=& (2+\frac{1}{L}) \theta(\frac{\l^n_\a}{n}) -
\frac{1}{L} \sum_{(m\beta)}\theta_{mn}(\l^n_\a
-\l^m_\b)+\theta_{mn}(\l^n_\a +\l^m_\b)\nn
&& + \frac{1}{L} \sum_{\gamma=1}^{N_h}\theta(\frac{\l^n_\a-\1l_\g}{n})
+\theta(\frac{\l^n_\a+\1l_\g}{n})+\frac{1}{L}\kappa^{(n)}_{ij}(\l^n_\a)\
, \a=1\ldots M_n\nn
\frac{2\pi}{L} J_\g &=& \frac{1}{L}\sum_{(n\alpha)}\theta(\1l_\g-\l^n_\a)
+\theta(\1l_\g+\l^n_\a)+ \frac{1}{L}\omega_{ij}(\1l_\g)\ , \g=1\ldots
M^{(1)} 
\label{baelog}
\eea
where $I^n_\a$ and $J_\g$ are integer numbers,
$\theta(x)=2\arctan(2x)$, 
\be
\theta_{n,m}(x) = (1-\delta_{m,n})\theta ({x\over{|n-m|}}) +
2\ \theta ({x\over{|n-m|+2}})+\dots +2\ \theta ({x\over{n+m-2}}) +
\theta ({x\over{n+m}})\ ,
\ee
and the boundary contributions are given by
\bea
\kappa^{(n)}_{ab}(\l)&=&\sum_{l=1}^n\theta(\frac{\l}{n+1-2l-S_1})\
,\quad \omega_{ab}(\l)= \theta(\frac{\l}{-S_L})\ ,\nn
\kappa^{(n)}_{ba}(\l)&=&\sum_{l=1}^n\theta(\frac{\l}{n+1-2l-S_L})\
,\quad \omega_{ba}(\l)= \theta(\frac{\l}{-S_1})\ ,\nn
\kappa^{(n)}_{bb}(\l)&=&\kappa^{(n)}_{ba}(\l)+\kappa^{(n)}_{ab}(\l)
,\quad \omega_{bb}(\l)= 0\ ,\nn
\kappa^{(n)}_{aa}(\l)&=&0,\quad \omega_{aa}(\l)=
\omega_{ba}(\l)+\omega_{ab}(\l)\ .
\label{bc}
\eea
The ranges of integers $I^{n}_\a$ and $J_\g$ are 
\be
I^n_\a=1,2,\ldots L+M_n-2\sum_{m=1}^\infty {\rm min}\{m,n\}M_m
+N_h\ ,\quad J_\g=1,2,\ldots N_\da+N_h-1\ .
\label{int}
\ee
There are two differences as compared to the case of periodic boundary
conditions \cite{fk,suth,schlott,bcfh,bares1,bares2,sarkar,ektj}: first
there are additional $\frac{1}{L}$ terms \r{bc}, and secondly the
integers $I^n_\a$ and $J_\g$ take different values.
The allowed range of the integers $I^n_\a$ and $J_\g$ reflects the
fact that all solutions of \r{bae} with one or more roots $\l_j$ or
$\1l_k$ having vanishing real parts must be excluded as they lead to
vanishing wave-functions. This restriction leads to constraints on the
allowed values of the integers $I^n_\a$ and $J_\g$: the $I^n_\a$ range
from $1$ to $L+M_n-2\sum_{m=1}^\infty {\rm min}\{m,n\} M_m +N_h$, the
solution with $I^n_\a=0$ being excluded. Similarly $J_\g$ range from
$1$ to $N_\da+N_h-1$ and $0$ is again excluded. 

For zero boundary fields ($\kappa^{(n)}_{ij}=0$, $\omega_{ij}=0$) we
can costruct a complete set of $3^L$ states from the Bethe Ansatz
states defined in the above way:
the model \r{hamil} with vanishing boundary fields is
$sl(1|2)$-invariant and all Bethe states are highest weight states of
$sl(1|2)$ \cite{fk2,gr}. Additional linearly independent eigenstates
of \r{hamil} can be constructed by acting with the $sl(1|2)$ lowering
operators on the highest-weight states. The total number of states
obtained in this way is $3^L$ as can be proved in the same way
as for the periodic $t$-$J$ chain \cite{fk} (the necessary
combinatorics are identical). Thus we obtain a complete set of
eigenstates of \r{hamil}.

For nonvanishing boundary fields the situation is more complicated as
the $sl(1|2)$ symmetry is broken by the boundary conditions. Therefore
we cannot use the symmetry generators to construct additional states
from the Bethe Ansatz states and are left with the a priori incomplete
set of eigenstates given by \r{baelog} and \r{int}. For the present
purposes this is inessential: the ground state is always a Bethe
Ansatz state, as are the states needed to extract the boundary
S-matrices. Ground state and excitations can be constructed from
\r{baelog} in a standard way (see {\sl e.g.} \cite{vladb}). The ground
state is obtained by filling all allowed vacancies of integers
$I^1_\a$ and $J_\g$ up to maximal values $I_{\rm max}$ and $J_{\rm
max}$, which corresponds to filling two Fermi seas of rapidities
$\l^1_\a$ between $0$ and $A$ and $\l^{(1)}_\g$ between $0$ and
$B$. The actual values of $A$, $B$ (and thus $I_{\rm max}$ and $J_{\rm
max}$) depend on $H$ and $\mu$ and are determined below. We are
interested in the case of a small magnetic field $H$ and a close to
half-filled band ($\mu\approx 2\ln(2)$), for which $A\gg 1$ and $B\ll
1$. As is shown in Appendix A the ground state energy per site (for
the four possible sets of boundary fields) below half-filling is given
by
\bea
\frac{E-\mu N_e -HS^z}{L}&=& \eps_c(0)-2\mu+\frac{1}{4\pi
L}\int_{-A}^A d\l\ \eps_s(\l) \kappa_{ij}^\prime(\l)
+ \frac{1}{4\pi L}\int_{-B}^B d\l\ \eps_c(\l)
\omega_{ij}^\prime(\l)\nn 
&&-\frac{1}{2L} [\eps_s(0)+\mu-\frac{H}{2}-2E_{ij}]+o(L^{-1}) \ ,
\label{E}
\eea
where $i,j=a,b$ and where the dressed energies $\eps_c(\l)$ and
$\eps_s(\l)$ \footnote{These can be shown to be (minus) the energies
of the order one contributions to the elementary charge and spin
excitations \cite{bares2}.} are given in terms of the coupled integral
equations  
\bea
\eps_s(\l)&=& -2\pi a_1(\l)+H-\int_{-A}^{A}d\mu\ a_2(\l-\mu)\ \eps_s(\mu)
+ \int_{-B}^B d\mu\ a_1(\l-\mu)\ \eps_c(\mu)\ ,\nn
\eps_c(\l)&=& \mu-\frac{H}{2}+\int_{-A}^{A}d\mu\
a_1(\l-\mu)\ \eps_s(\mu)\ .
\label{dresseden}
\eea
Here $a_n(\l) = \frac{1}{2\pi} \frac{n}{\l^2+\frac{n^2}{4}}$. For
later use we define
\be
G_x(\l)=\frac{1}{2\pi}\int_{-\infty}^\infty d\omega\ \exp({-i\omega\l})
\frac{\exp(-x|\frac{\omega}{2}|)}{2\cosh(\frac{\omega}{2})}=
\frac{1}{\pi} {\rm Re}\left(\beta(\frac{1+x}{2}+i\l)\right)\ ,
\ee
where $x$ is real and where
$\beta(z)=\frac{1}{2}\left[\psi(\frac{1+z}{2})-\psi(\frac{z}{2})\right]$.
Here $\psi(z)$ is the digamma function. The asymptotic behaviour of
$G_x(\l)$ for large $l\gg 1$ and $\l\gg x$ is
\be
G_x(\l)\sim \frac{1}{4\pi}\frac{x}{\l^2}+{\cal O}(\l^{-4})\ .
\label{asy}
\ee
Below we will also need the small-$\l$ asymptotics of $G_1(\l)$, which
is given by
\be
G_1(\l) = \frac{1}{2\pi}[2\ln(2)+2\sum_{n=1}^\infty (-1)^n
(1-2^{-2n})\zeta(2n+1) \l^{2n}]\ ,\ |\l|<1\ .
\ee
\section{\sc Wiener-Hopf Analysis for the Dressed Energies}

In this section we analyze the coupled integral equations
\r{dresseden} by means of Wiener-Hopf techniques \cite{yaya} (for
detailed expositions see {\sl e.g.} \cite{kondo,nep1}). As \r{dresseden}
are similar to the analogous equations for the densities in the
periodic $t$-$J$ chain the necessary steps are the same as in
\cite{schlott}. However as we will need more explicit answers than are
given in \cite{schlott} for determining the boundary contribution to
the ground state energy we  briefly summarize the most important steps
below. After Fourier-transforming, the first equation of \r{dresseden}
%%
%%eqns \r{dresseden} are rewritten as 
%%\bea
%%\eps_s(\l)&=& -2\pi
%%G_0(\l)+\frac{H}{2}+\left\{\int_{A}^{\infty}+\int_{-\infty}^{-A} \right\}
%%d\nu\ G_1(\l-\nu)\ \eps_s(\nu) + \int_{-B}^B d\nu\ G_0(\l-\nu)\
%%\eps_c(\nu)\ ,\nn 
%%\eps_c(\l)&=& \mu-2\pi
%%G_1(\l)-\left\{\int_{A}^{\infty}+\int_{-\infty}^{-A}\right\} 
%%d\nu\ G_0(\l-\nu)\ \eps_s(\nu)+ \int_{-B}^B d\nu\ G_1(\l-\nu)\
%%\eps_c(\nu)\ .
%%\label{wh0}
%%\eea
%%The first of these equations 
can be turned into a Wiener-Hopf equation for $y(\l)=\eps_s(\l+A)$
\be
y(\l)= -2\pi G_0(\l+A)+\frac{H}{2}+\int_{0}^{\infty}d\nu\ [G_1(\l-\nu)
+G_1(\l+\nu+2A)]\ y(\nu) + C G_0(\l+A)\ ,
\label{wh}
\ee
where $C=\int_{-B}^Bd\nu\ \exp(\pi\nu)\ \eps_c(\nu)$. Here we have
used the fact that $A\gg 1$ and $B\ll 1$ to approximate 
$ \int_{-B}^B d\nu\ G_0(\l-\nu+A)\ \eps_c(\nu)\approx
G_0(\l+A)\int_{-B}^Bd\nu\ \exp(\pi\nu)\ \eps_c(\nu)$.
The quantity $C$ is determined self-consistently below.
Eqn \r{wh} can now be solved by iteration
$y(\l)=y_1(\l)+y_2(\l)+\ldots$, where
\bea
y_1(\l)&=& -2\pi G_0(\l+A)+\frac{H}{2}+\int_{0}^{\infty}d\nu\ G_1(\l-\nu)
\ y_1(\nu) + C G_0(\l+A)\ ,\\
y_2(\l)&=& \int_{0}^{\infty}d\nu\ G_1(\l+\nu+2A)\ y_1(\nu) +
\int_{0}^{\infty}d\nu\ G_1(\l-\nu)\ y_2(\nu)\ .
\label{y}
\eea
These equations can be solved in a standard way through a Wiener-Hopf
factorization. The result for $y_1$ is obtained in complete analogy with
{\sl e.g.} the Appendix of \cite{holger} 
\be
\widetilde{y}^+_1(\omega)=G^+(\o)\left\lbrace\frac{iH
G^-(0)}{2}\frac{1}{\o+i0} -i (2\pi-C)G^-(-i\pi)\frac{\exp(-\pi
A)}{\o+i\pi}\right\rbrace + {\cal O}(\exp(-2\pi A))\ .
\label{y1}
\ee
Here the Fourier transform
$\widetilde{y}^+_1(\omega)=\int_0^\infty d\l\ y(\l) \exp(i\l\o)$
is analytic in the upper half-plane and $G^\pm(\o)$ are analytic
functions in the upper/lower half-plane factorizing the kernel
$1+\exp(-|\o|)= G^+(\o)G^-(\o)$
\be
G^+(\o)=G^-(-\o)=\frac{\sqrt{2\pi}}{\Gamma(\frac{1}{2}-\frac{i\o}{2\pi})}
\left( \frac{-i\o}{2\pi}\right)^\frac{-i\o}{2\pi}
\exp(\frac{i\o}{2\pi})\ .
\ee
The equation for $y_2(\l)$ is more difficult to solve. They key is to
use the fact that $\l+\l^\prime +2A\gg 1$. Using the asymptotic
behaviour \r{asy} of $G_1(\l)$ in the expression for the driving term
$D(\l)=\int_0^\infty d\l^\prime\ G_1(\l+\l^\prime +2A) y_1(\l^\prime)$
and then performing a Laplace transformation we obtain
\be
D(\l)\sim \frac{1}{4\pi}\int_0^\infty dx\ \exp(-2Ax) \exp(-|\l|x)
{\widetilde y}_1^+(ix) [x+\frac{x^3}{12}+\ldots]\ ,
\label{dri}
\ee 
where the expansion in $x$ corresponds to the asymptotic expansion of
$G_1(\l+\l^\prime+2A)$. It is clear that due to the strongly decaying
factor $\exp(-2Ax)$ the leading contribution to the integral comes
from the small-$x$ region. Inserting the expression \r{dri} for the
driving term into \r{y} and then following through the same steps as
in the analysis for $y_1$ we arrive at
\bea
{\widetilde y_2}^+(\o)&\sim& G^+(\o) \frac{i}{4\pi}\int_0^\infty dx\
\exp(-2Ax) [x+\frac{x^3}{12}+\ldots] \frac{G^+(ix)
{\widetilde y_1}^+(ix)}{\o+ix}\nn
&\sim& G^+(\o) \frac{iH\sqrt{2}}{4\pi}\int_0^\infty dx\
\exp(-2Ax) \frac{1 +\frac{x}{2\pi}\ln(x)+\ldots}{\o+ix}.
\label{y2}
\eea
By means of a similar analysis further corrections to $y(\l)$ can be
determined. As far as the physical quantities determined below are
concerned $y_3$, $y_4$ {\sl etc} give rise to contributions much
smaller than those due to $y_1$ and $y_2$.
We are now in a position to determine the limit of integration $A$ as
a function of the magnetic field. By definition $\eps_s(\pm A)=0=y(0)$,
which leads to
\be
A= -\frac{\ln(H)}{\pi}+\frac{\ln(\sqrt{\frac{2\pi}{e}} 
(2\pi-C))}{\pi}+\frac{1}{4\pi\ln(H)}+\ldots\ .
\ee
Using \r{y1} and \r{y2} we can now solve the integral equation
\r{dresseden} for $\eps_c(\l)$ 
\bea
\eps_c(\l)
%&=& \mu -2\pi G_1(\l) -\int_0^\infty d\nu\
%[G_0(\l-\nu-A)+G_0(\l+\nu+A)] y(\nu)\nn
%&& +\int_{-B}^Bd\nu\ G_1(\l-\nu)\ \eps_c(\nu)\nn
&=& \eps_0(\l)+\int_{-B}^Bd\nu [G_1(\l-\nu)+
\frac{2a\cosh(\pi\l)}{2\pi}\exp(\pi\nu)] \eps_c(\nu)\ ,
\label{epsc}
\eea
where $a=\frac{\pi}{e}\exp(-2\pi A)$, $\eps_0(\l)=\mu -2\pi G_1(\l)
-2a\cosh(\pi\l)$, and where we have neglected terms of order
$o(\exp(-2\pi A))$. Here the term proportional to $a$ originates in
the $C$-term in \r{y1}. Equation \r{epsc} can now be solved by
iteration as $B\ll 1$ (corresponding to $|\mb|=|2\ln(2)-\mu|\ll 1$)
with the result 
\bea
\eps_c(\l) &=& \mu -(2\pi+g) G_1(\l)-2a\cosh(\pi\l)
+{O}(\mb a) + {\cal O}(\mb^2) + {\cal O}(a^2)\ ,
\eea
where
\be
a= \frac{H^2}{8\pi^2} (1-\frac{1}{2\ln(H)}+\ldots )\ ,\qquad
g=\frac{8}{3}\frac{1}{\sqrt{6\zeta(3)}}(\mb+2a)^\frac{3}{2}\ .
%%+\frac{64\ln(2)}{9\pi \zeta(3)} \mb a \ ,
\ee
The boundary of integration $B$ defined {\sl via} $\eps_c(\pm B)=0$ in
this order is given by 
%%\bea
%%B^2&=&{B_0^2+\frac{8\ln(2)}{9\pi\zeta(3)}\sqrt{\frac{2}{3\zeta(3)}}(\mb
%%+2a)^\frac{3}{2} +\frac{128 (\ln(2))^2}{27\pi^2(\zeta(3))^2}\mb a}\ ,\nn
%%B_0^2&=&\frac{2}{3\zeta(3)}\left[\mb+2a+\mb a
%%\left(\frac{2\pi^2}{6\zeta(3)}+
%%\frac{10\zeta(5)}{3\zeta(3)^2}\right)\right]\ .
%%\eea
\be
B^2=\frac{2}{3\zeta(3)}\left[\mb+2a+\frac{8\ln(2)}{3\pi}
\frac{1}{\sqrt{6\zeta(3)}}(\mb+2a)^{\frac{3}{2}}\right]
\label{B2}
\ee
$C$ is determined self-consistently to be $-2\zeta(3)B^3$. The higher
order (in $B$) contributions to $\eps_c$ and $B$ do not contribute to
the singularities in the thermodynamic quantities and therefore have
been dropped. 

\section{\sc Ground State Properties}

We are now in a position to determine bulk and boundary contributions
to the energy \r{E}. The bulk energy per site is found to be 
\be
e_{bulk} = -[\mu+2\ln(2)+2a+2\ln(2)\frac{\zeta(3)}{\pi}B^3]\ ,
\ee
from which we can determine the leading contributions to the
zero-temperature magnetization per site, magnetic susceptibility,
density and compressibility close to half-filling in a weak magnetic
field 
\bea
m_{bulk}&=&-\frac{\partial e_{bulk}}{\partial H} =
\frac{H}{2\pi^2}(1-\frac{1}{2\ln(H)})
\left(1+\frac{\ln(2)}{\pi}\sqrt{\frac{8(\mb+2a)}{3\zeta(3)}}\right)
+\ldots\ ,\nn
\chi_{H,bulk}&=& \frac{1}{2\pi^2}(1-\frac{1}{2\ln(H)})
+\frac{2\ln(2)}{\pi}\frac{H^2}{4\pi^4}(1-\frac{1}{\ln(H)})
\frac{1}{\sqrt{6\zeta(3)(\mb+2a)}}+\ldots\ ,\nn
D_{bulk}&=&-\frac{\partial e_{bulk}}{\partial \mu} = 1 -
\frac{2\ln(2)}{\pi}\sqrt{\frac{2(\mb+2a)}{3\zeta(3)}}+\ldots\ ,\nn
\chi_{c,bulk}&=&\frac{1}{D_{bulk}^2}\frac{\partial D_{bulk}}{\partial\mu} =
\frac{2\ln(2)}{\pi}\frac{1}{\sqrt{6\zeta(3)[\mb+2a]}}+\ldots\ ,
\eea
in agreement with the expressions for periodic boundary conditions
\cite{schlott,bcfh}. We note that both the magnetic susceptibility
and the compressibility diverge when we approach half-filling.

Contributions to the surface energy, {\sl i.e.} all terms
proportional to $L^{-1}$ in \r{E}, can be divided into
boundary-field dependent ones $E^{(\a\b)}$ and contributions
due to the ``geometry'' {\sl i.e.} openess of the chain $E^0$ so that
we can write for the four permitted sets of boundary conditions
\be
E_{boundary} = E^0+E^{(\a\b)}\ ,\qquad\a,\b=a,b\quad .
\ee
The boundary field independent contributions are easily determined
\be
E^0= -\frac{1}{2}\left\lbrace-\frac{H}{2\ln(H)}
\left(1+\frac{\ln(\ln(H))}{2\ln(H)}\right)
+\mu-\pi-\sqrt{\frac{8}{27\zeta(3)}}(\mb+2a)^\frac{3}{2}+
\ldots\right\rbrace . 
\ee
We note that for zero bulk magnetic field $H=0$ and half-filling
$\mu=2\ln 2$ we obtain $E^0=\frac{\pi}{2}-\ln 2$, which is the
correct result for the surface energy of the open XXX Heisenberg chain
\cite{xxz1,gmn}. By differentiating the surface energy with respect to
the thermodynamic parameters $H$ and $\mu$ we can evaluate the surface
contributions to particle number and magnetization in analogy with
{\sl e.g.} the treatment of the Kondo model \cite{kondo} (see also
\cite{tsv}). It is reasonable to assume that these contributions are
concentrated in the boundary regions, {\sl e.g.} we interpret the
surface contribution to the particle number to lead to a
depletion/increase of electrons in the ``vicinity'' of the boundaries.

The leading contributions to the boundary magnetization, particle
number and susceptibilities due to $E^0$ are
\bea
M^0&=&-\frac{1}{4\ln(H)}
\left(1+\frac{\ln(\ln(H))}{2\ln(H)}\right)+\ldots\ ,\quad
\chi^0_H=\frac{1}{4H(\ln(H))^2}
\left(1+\frac{\ln(\ln(H))}{2\ln(H)}\right)+\ldots\ ,\nn
N^0&=& \frac{1}{2}\left(1 +
\sqrt{\frac{2(\mb+2a)}{3\zeta(3)}}\right),\quad
\chi^0_c=-\frac{4\ln 2+\pi}{2\pi\sqrt{6\zeta(3)(\mb+2a)}}\ .
\eea
We first note that boundary region exhibits a stronger magnetization
as compared to the bulk, {\sl i.e.}
$\frac{M^0}{m_{bulk}}=-\frac{\pi^2}{2 H\ln H}$, which is much larger
than one for the small magnetic fields considered here. 
The boundary magnetic susceptibility is seen to diverge for
$H\rightarrow 0$. Following \cite{tsv} we interpret this as an
indication for the presence of a magnetic bound state in the boundary
region. The magnetic behaviour is similar to the one for the XXZ spin
chain with an open boundary studied in \cite{tsv}. 

The leading contribution to the boundary particle number is
$\frac{1}{2}$. Because of the constraint of at most single occupancy
at any site this increase in particle number (recall that we are very
close to half-filling) must be spread out over large regions
neighbouring the boundaries. This indicates that boundary effects
spread deeply into the bulk. The boundary compressibility for zero
boundary fields is seen to be negative and to diverge as we approach
half-filling. The type of singularity is the same as for the bulk. 
Combining the results for magnetization and particle number we see
that there is a tendency for spin-up electrons to get pushed towards
the boundary.

The leading order boundary-field dependent contributions $E^{(\a\b)}$
are expressed (see \r{E}) in terms of the quantities 
\be
\epsilon_b(S)= \int_{-A}^Ad\nu\ a_S(\nu)\ \eps_s(\nu)\ ,\quad
\epsilon_a(S)= \int_{-B}^Bd\nu\ a_S(\nu)\ \eps_c(\nu)\ .
\ee
where according to \r{E} for the four possible sets of boundary
conditions 
\be
E^{(\a\b)}=
\frac{1}{2}\left[\epsilon_\b(-S_1)+\epsilon_\a(-S_L)\right]+E_{\a\b}
,\quad \a,\b=a,b\ .
\ee

The leading contribution to the quantity $\epsilon_a(S)$ can be easily
determined for the case $S\gg B$, in which we can expand
$a_S(\nu)$ in a power series in $\nu$ and then perform the elementary
integrations using the expression \r{epsc} for $\eps_c(\nu)$. 
For $S\ll B$ we instead expand $\eps_c(\nu)$ in an infinite power
series (using that $G_1(\nu)$ is a smooth function around zero), then
perform the integrations, resum the result, and then retain only the
leading terms. This results in
\be
\epsilon_a(S)= \cases{-\frac{4\zeta(3)B^3}{\pi}\frac{1}{S}+\ldots &{\rm if}
$S\gg B$\cr
\eps_c(0)+\frac{2S}{\pi B}(\mb+2a+\ldots)+\ldots& {\rm if} $S \ll
B$\ .\cr}
\label{epsa}
\ee
The analogous computations for $\epsilon_b(S)$ are more involved, so
that we give a brief summary of the necessary steps in Appendix B. We
find the following result for the leading behaviour 
\be
\epsilon_b(S)=-G_S(0) (2\pi+2\zeta(3) B^3)+\frac{H}{2}+
\cases{\frac{S-1}{2}\frac{H}{\ln(H)}+\ldots & {\rm if} $S\ll A$\cr
-\frac{H}{2}-\frac{2}{\pi^2S}H\ln(H)+\ldots &
{\rm if} $S\gg A$\cr}\ .
\label{epsb}
\ee

\subsection{\sc Contributions due to small boundary fields of type $a$} 

The contribution to the boundary energy is given by
$\frac{1}{2}\epsilon_a(\frac{2}{h}-2)$ with $\frac{2}{h}-2\gg
B=\sqrt{\frac{2(\mb+2a)}{3\zeta(3)}}$. (This defines what we mean with
``small'' boundary field). Here $h$ is a boundary chemical
potential. We define the quantity
$\s=\frac{h}{1-h}\sqrt{\frac{\mb+2a}{6\zeta(3)}}$, which in the
present case is much smaller than one. We obtain the following
contributions to boundary magnetization/particle number and
susceptibilities    
\bea
M^a&=&\frac{H}{\pi^3} (1-\frac{1}{2\ln(H)})\s\ ,\qquad
N^a=-\frac{2}{\pi}\s\ ,\nn
\chi_H^a&=&\frac{1}{\pi^3}(1-\frac{1}{2\ln(H)})\s+\frac{H^2}{4\pi^5}
(1-\frac{1}{2\ln(H)})\s(\mb+2a)^{-1}\ ,\nn
\chi^a_c&=&\frac{\s}{\pi}(\mb+2a)^{-1}\ .
\eea
We see that a small boundary chemical potential leads essentially to
the same type of divergences as are present in the bulk.
As expected electrons get pushed away from the boundary although the
effect is small. By differentiation with respect to the boundary
chemical potential we can evaluate the average number of electrons at
the boundary site 
\be
\langle n_e \rangle =1-\frac{2\zeta(3)}{\pi}\frac{B^3}{(1-h)^2}\ ,\quad
h\rightarrow 0\ ,
\ee
where $B$ is given by \r{B2}. We see that the electron number is
larger than the bulk value. This is consistent with the above
observation that an open boundary without field leads to an increase
in the elctron density in the boundary region.

\subsection{\sc Contributions due to large boundary fields of type $a$} 

Here the boundary chemical potential is taken large, by which we mean
that $0<\frac{2}{h}-2\ll B$. We again use the notation
$\s=\frac{h}{1-h}\sqrt{\frac{\mb+2a}{6\zeta(3)}}$, but now $\s\gg
1$. We find 
\bea
M^a&=&\frac{H}{4\pi^2}(1-\frac{1}{2\ln(H)})(1-\frac{1}{\pi\s})\
,\quad 
N^a=-\frac{1}{2}(1+\frac{2\ln(2)}{\pi}
\sqrt{\frac{2(\mb+2a)}{3\zeta(3)}}) +\frac{1}{2\pi\s}\ ,\nn
\chi^a_H&=& \frac{1}{4\pi^2}(1-\frac{1}{2\ln(H)})(1-\frac{1}{\pi\s})+
\frac{1}{\pi\s}\frac{H^2}{(2\pi)^4}(1-\frac{1}{\ln(H)})(\mb+2a)^{-1}+\ldots\
,\nn 
\chi^a_c&=&\frac{3\ln(2)}{\pi\sqrt{6\zeta(3)(\mb+2a)}}
+\frac{1}{4\pi\s}(\mb+2a)^{-1}\ .
\eea
The magnetization is again proportional to $H$ and the magnetic
susceptibility can only diverge at half-filling.
The large boundary field yields however a contribution of
$-\frac{1}{2}$ to the boundary particle number, which indicates a strong
depletion of electrons in the boundary region. This is in accordance
with the expectation based on a naive analysis of the hamiltonian
\r{hamil} that large boundary chemical potentials (with our choice of
sign in \r{hamil}) should favour the presence of holes in the boundary
region. The compressibility exhibits a stronger divergence than the
bulk if we approach half-filling keeping $\s$ fixed.
The average electron number at the boundary site is found to be 
\be
\langle n_e \rangle =1-\frac{2}{\pi h^2}\sqrt{6\zeta(3)(\mb+2a)}\
,\quad h\rightarrow 1,
\ee
which is less than the bulk value.

\subsection{\sc Contributions due to large boundary fields of type $b$} 

Let us first consider the region where $h|\ln H|\gg 2\pi$, which
corresponds to the case $S\ll A$. By straightforward differentiation
we find
\bea
M^b&=&-\frac{1}{4}+\frac{h-1}{2h\ln H}+
\frac{HB}{4\pi^3}\left[\psi(\frac{1+h}{2h})-\psi(\frac{1}{2h})\right]\nn
\chi_H^b&=&\frac{1-h}{2hH\ln^2H}+\frac{1}{4\pi^3}\left[\psi(\frac{1+h}{2h})
-\psi(\frac{1}{2h})\right]\left(B+\frac{H^2}{6\pi^2\zeta(3)B}\right)\nn
N^b&=&
-\frac{1}{2\pi}\left[\psi(\frac{1+h}{2h})-\psi(\frac{1}{2h})\right]
\sqrt{\frac{2(\mb+2a)}{3\zeta(3)}}\nn
\chi^b_c&=&
\frac{1}{2\pi}\left[\psi(\frac{1+h}{2h})-\psi(\frac{1}{2h})\right]
\frac{1}{\sqrt{6\zeta(3)(\mb+2a)}}\ .
\label{blarge}
\eea
Thus the boundary field contributes to the singularity in the magnetic
susceptibility for large boundary fields $h$. Furthermore there is a
constant contribution $-\frac{1}{4}$ to the boundary magnetization,
which indicates a surplus of spin-down electrons in the boundary
region. The negative sign stems from the fact that the boundary field
is antiparallel to the bulk magnetic field. The boundary particle
number contribution is always small (and leads to a depletion of the
electron number in the boundary region) and the compressibility
exhibits the same type of divergence as the bulk. By differentiating
with respect to the boundary field we can calculate the $\langle
S^z-\frac{n^h}{2}\rangle$ at site $1$ ($L$) for $ab$ ($ba$) boundary
conditions. The result is 
\be
\langle S^z-\frac{n^h}{2}\rangle=\frac{1}{2}-\int_0^\infty d\omega\
\omega \frac{\exp(-\omega)}{1+\exp(-h\omega)}-\frac{H}{2h^2\ln H}\ . 
\ee
This is always finite in the range of $h$ considered. To get a rough
idea of the contribution of the integral we note its respective values
for $h=1$ and $h=2$, which are $\frac{\pi^2}{12}$ and $0.91597$
(Catalan's constant). The contribution proportional to $H$ is always
small. The susceptibility is finite, which means that the spins and
charges at the boundary itself do not contribute to the bound state
responsible for the singularities in the susceptibilities.

\subsection{\sc Contributions due to small boundary fields of type $b$} 

Let us now turn to the region $h|\ln H|\ll 2\pi$ of vanishingly small
boundary fields. We find
\bea
M^b&=&\frac{h\ln(H)}{2\pi^2}\ ,\qquad \chi^b_H=\frac{h}{2\pi^2H}\ ,\nn
N^b&=&-\frac{h}{2\pi}\sqrt{\frac{2}{3\zeta(3)}(\mb+2a)}\ ,\qquad
\chi^b_c=\frac{h}{2\pi\sqrt{6\zeta(3)(\mb+2a)}}\ .
\eea
Note that we cannot take $H\rightarrow 0$ without taking the boundary
field to zero first. Thus the magnetization is always small. However
the boundary magnetic susceptibility may or may not diverge for
$H\rightarrow 0$, depending on how fast we take take the boundary
field to zero as compared to the bulk field. 
The result for $\langle S^z-\frac{n^h}{2}\rangle$ is now found to be
\be
\langle S^z-\frac{n^h}{2}\rangle = -\frac{H\ln(H)}{2\pi^2}\ .
\ee
This is again small and vanishes for half-filling and zero bulk
field in accordance with \cite{tsv}. The corresponding susceptibility
is again finite and the spins/charges at the boundary site do not
contribute to the bound state.

\section{\sc Boundary S-Matrix}

In this section we study the effects of the boundary on gapless
excitations. For simplicity we set the bulk magnetic field $H$ to
zero. As the elementary excitations in the bulk are the same for the
periodic and the open chain we begin by giving a thorough discussion
of the interpretation of the spectrum in terms of elementary
excitations for the periodic system. After reviewing the known
results of \cite{bares2,fk} we present a conjecture concerning
the $sl(1|2)$ descendants of the holon and spinon states obtained from
the Bethe Ansatz.

There are two kinds of gapless elementary excitations in the
supersymmetric $t$-$J$ model, associated with spin- and charge degrees
of freedom respectively. For the periodic chain they have been
extensively studied in \cite{bares1,bares2} (see below for a
discussion of the $sl(1|2)$ structure of the excitations). The spin
excitations are called spinons and carry spin $\pm \frac{1}{2}$ and
zero charge. They are very similar to the spin-waves in the Heisenberg
XXX chain \cite{faddtak}. The charge excitations are called holons and
antiholons, carry zero spin and charge $\mp e$. Holons correspond to
``particles'' in the charge Fermi sea of $\l^{(1)}$'s and are thus
associated with a physical hole, whereas antiholons correspond to
``holes'' (unoccupied $\l^{(1)}$'s) in the charge Fermi sea. 
At half-filling only holons can be excited as the charge Fermi sea is
completely empty. The excitation energies are given by
$\eps_{s,c}(\l)$ \r{dresseden}, whereas their physical momenta (for
the periodic chain) are given in terms of the quantities
${\tt p}_{s,c}(\l)$, which are solutions of the integral equations
\bea
{\tt p}_s(\l)&=& -\theta(\l)-\int_{-A}^Ad\nu\ a_2(\l-\nu) {\tt p}_s(\nu)
+ \int_{-B}^B d\nu\ a_1(\l-\nu)\ {\tt p}_c(\nu)\ ,\nn 
{\tt p}_c(\l)&=& \int_{-A}^A d\nu\ a_1(\l-\nu)\ {\tt p}_s(\nu).
\label{mtm}
\eea
The mometum of {\sl e.g.} a holon-antiholon excitation is given by
$P_{c{\bar c}}={\tt p}_c(\La^p)-{\tt p}_c(\La^h)$ where $\La^p$ and
$\La^h$ are the spectral parameters of the holon and antiholon
respectively. We thus would define the physical holon momentum as
$p_c(\La^p) = {\tt p}_c(\La^p)-{\tt p}_c(B)$.
At half-filling the spinon ($p_s$) and holon ($p_c$) momenta are given
by 
\bea
p_s(\l)&=& 2\arctan(\exp(\pi\l))-\pi\ ,\nn
p_c(\l)&=& \frac{\pi}{2}+i\ \ln\left(
\frac{\Gamma(\frac{1-i\l}{2})}{\Gamma(\frac{1+i\l}{2})}
\frac{\Gamma(1+\frac{i\l}{2})}{\Gamma(1-\frac{i\l}{2})}\right).
\label{mtmhf}
\eea
The ``order one'' contributions to the spinon and holon energies at
half-filling take the simple form $\eps_s(\l) = 2\pi G_0(\l)$ and
$\eps_c(\l) = 2\ln(2)-2\pi G_1(\l)$. The spinon dispersion
is thus of the Faddeev-Takhtajan form $\eps_s(p) = \pi\sin(p)$.
The holon dispersion cannot be written in closed form so easily.

So far we have not discussed the role played by the $sl(1|2)$ symmetry
in the classification of eigenstates. As was shown in \cite{fk} all
eigenstates of the hamiltonian obtained from the Bethe Ansatz are
highest weight states of the $sl(1|2)$ symmetry algebra of the
model. Additional eigenstates of the hamiltonian can be obtained by
acting with the $sl(1|2)$ generators (recall that we are still
discussing the periodic chain). This leads to the picture put forward
in \cite{fk} for the structure of excitations over the
antiferromagnetic ground state, {\sl i.e.} spinon and holon/antiholon
excitations are really only the highest weight states of $sl(1|2)$
multiplets. However all the additional excitations 
constructed by acting with the symmetry generators with the exception
of the spin-$SU(2)$ descendants can be easily shown to have a gap
proportional to the chemical potential (see also \cite{ek}). Therefore
we can obtain a {\sl complete} set of {\sl gapless} excitations by
taking into account spinon and holon states plus their spin-$SU(2)$
descendants.  
However, one can furthermore argue that in the thermodynamic limit
({\sl i.e.} if we neglect all finite-size corrections) actually all
the gapped $sl(1|2)$ descendants obtained by acting with the symmetry
generators can be viewed as being incorporated in multiparticle spinon
and holon/antiholon excitations. Let us first 
discuss the situation for a large finite chain. Here a complete set of
states is given by first finding all sets of spectral parameters
solving the Bethe equations \r{bae}. Each such solution yields the
wave-function of a highest-weight state of $sl(1|2)$ with a given
fixed momentum, and a complete set of states is obtained by taking
into account the $sl(1|2)$ descendants of the highest-weight state.
As we approach the thermodynamic limit we identify one-parameter
families of highest weight states as elementary excitations, the free
parameter being the rapidity (which is directly related to the
momentum) of the particle. Thus in the thermodynamic limit we are
interested in multiparameter families of excited states and the strict
counting of states possible in the finite volume loses its meaning.
This is the reason why the $sl(1|2)$ descendants of the spinon and
holon/antiholon excitations do not have to be taken into account
separately anymore in the thermodynamic limit. From the
Algebraic Bethe Ansatz construction we know that the symmetry
generators can be represented as the infinite spectral parameter
limits of the elements of the monodromy matrix \cite{fk,eks}. On the
level of the Bethe Ansatz states this means that the action of the
symmetry generators can be implemented by taking a spectral parameter
in an appropriate Bethe Ansatz state to infinity. If we therefore take
$k$ rapidities of an $n$ parameter family of excited (highest weight)
states to infinity we produce an $n-k$ parameter family of exited
states made of $sl(1|2)$ descendants! This means that the family of
$sl(1|2)$ descendant states does not have to be taken into account
separately anymore, as can equally well be interpreted as ``sitting on
the boundary'' of the $n$-parametric family of highest-weight states.
In this way we obtain an equivalent yet different
``quasiparticle-interpretation'' in the spirit of McCoy {\sl et. al.}
\cite{barry}.

As the simplest example let us consider the $sl(1|2)$ descendants of
the antiferromagnetic ground state at half-filling, which sits in a
multiplet of dimension four, the other three states being $Q_\sigma
|{\rm GS}\rangle$ ($\sigma=\up,\da$) with momentum zero and energy
$E=2\ln 2$ and $Q_\da Q_\up |{\rm GS}\rangle$ with momentum zero and
$E=4\ln 2$. The state $Q_\da |{\rm GS}\rangle$ can be obtained from
the spinon-holon scattering state (or more precisely the
two-parametric family of states) by taking the spectral parameters of
both the holon ($\La^p$) and the spinon ($\l^h$) to infinity: indeed
the quantum numbers match and $\eps_s(\l^h)+\eps_c(\La^p)\rightarrow
2\ln 2$ and $p_s(\l^h)+p_c(\La^p)\rightarrow 0$. The state $Q_\up
|{\rm GS}\rangle$ can then be obtained by acting with the spin raising
operator $S^\dagger$. Similarly the state $Q_\da Q_\up |{\rm
GS}\rangle$ is obtained from the (two-parametric) holon-holon
scattering state by again taking both spectral parameters to infinity.

On the basis of the above discussion we therefore propose that the
quasiparticle interpretation of the order one excitation spectrum in
terms of two spinons with spin $\pm\frac{1}{2}$ and holon and
antiholon excitations put forward in \cite{bares2} does actually already
incorporate the complete $sl(1|2)$ structure. For the half-filled band
it is straightforward to show \footnote{The computation is analogous
to the appendix of \cite{ekh2}.} by using the distribution of integers
\r{int} and the highest weight theorem that all gapless excitations
are scattering states of two spinons and one holon with the
superselection rule that the number of spinons plus the number of
holons is even. Thus the excitation spectrum of the half-filled
$t$-$J$ model is described by a $SU(2)\times U(1)$ scattering theory. 

The scattering matrix has been determined by means of Korepin's method
\cite{korepin,vladb} in \cite{bares1,bares2}. At half-filling the
spinon-spinon S-matrix $S(\l)$ and the spinon-holon ($sc$) and
holon-holon ($cc$) scattering phases are given by
\bea
S(\l) &=& i\frac{\Gamma(\frac{1-i\l}{2})}{\Gamma(\frac{1+i\l}{2})}
\frac{\Gamma(1+\frac{i\l}{2})}{\Gamma(1-\frac{i\l}{2})}
\left(\frac{\l}{\l-i}I - \frac{i}{\l-i} P\right)\ ,\nn
\exp(i\varphi_{sc}(\l))&=& -i \frac{1+ie^{\pi\l}}{1-ie^{\pi\l}}\ ,\qquad
\exp(i\varphi_{cc}(\l))=
\frac{\Gamma(\frac{1+i\l}{2})}{\Gamma(\frac{1-i\l}{2})}
\frac{\Gamma(1-\frac{i\l}{2})}{\Gamma(1+\frac{i\l}{2})}
\label{smhf}
\eea
where $I$ and $P$ are the $4\times 4$ identity and permutation
matrices respectively. Below half-filling the S-matrices are given in
terms of the solution of integral equations.
This concludes our discussion of the excitation spectrum of the
periodic $t$-$J$ model. For the case of the open chain the elementary
excitations are identical as are the bulk S-matrices. What remains in
order to completely specify the scattering of elementary excitations
is to determine the phase-shifts acquired by the spinons and holons
when reflecting from one of the boundaries. Note that due to the
particular form of the boundary interactions (no particle production
or transmutation) it is clear that the boundary S-matrices for holons
and spinons are diagonal and thus reduce to phase-shifts for the
physical states.

\subsection{\sc Boundary S-matrix for the exactly Half-Filled Band}

For the case of the exactly half-filled band the boundary S-matrix can
be determined by the method introduced by the Miami group in
\cite{gmn} for the case of the spin $\frac{1}{2}$ XXX Heisenberg
chain, which generalizes the methods of Korepin \cite{korepin} and
Andrei {\sl et al} \cite{al,ad}. An alternative method was introduced
in \cite{saleur} and can be seen to lead to the same results.
In order to determine the boundary phase shifts for spinons and holons
we will study the spinon-holon scattering state.
The method of \cite{gmn} is based on the following quantization
condition for factorized scattering of two particles with rapidities
$\l_{1,2}$ on a line of length $L$ (see also \cite{saleur})
\be
\exp(2iLp(\l_1))S(\l_1-\l_2)K_1(\l_1,h_1)S(\l_1+\l_2)
K_1(\l_1,h_L)=1\ ,
\label{qc}
\ee
where $p(\l)$ is the expression for the physical momentum of the
corresponding (infinite) periodic system, $S(\l)$ are the bulk
scattering matrices for scattering of particles $1$ and $2$, and
$K_1(\l,h)$ are the $K$-matrices describing scattering of particle $1$
with rapidity $\l$ off a boundary with boundary field $h$. For the
present case of spinon-holon scattering with boundary fields
preserving the two $U(1)$ symmetries of spinon and holon numbers this
condition turns into scalar equations for the scattering phases, which
after taking the logarithm read
\bea
2Lp_s(\l^h)+\varphi_{sc}(\l^h-\La^p)+\psi_s(\l^h,h_1)+\varphi(\l^h+\La^p)+
\psi_s(\l^h,h_L)=0\ {\rm mod}\ 2\pi\ ,\nn
2Lp_c(\La^p)+\varphi_{sc}(\La^p-\l^h)+\psi_c(\La^p,h_1)+\varphi(\La^p+\l^h)+
\psi_c(\La^p,h_L)=0\ {\rm mod}\ 2\pi\ .
\label{qchf2}
\eea
Here $\l^h$ and $\La^p$ are the rapidities of the spinon and holon
respectively. Comparing these conditions with certain quantities
(``counting functions'') that can be calculated from the Bethe Ansatz
solution one can then read off the boundary phase-shifts $\psi_{s,c}$
\cite{gmn}. Let us now discuss this program for the spinon-holon
scattering state characterized by choosing $M_1=\frac{L}{2}, N_h=1$ in
the Bethe equations \r{baelog}. There are $\frac{L}{2}+1$ vacancies
for the integers $I^1_\a$ and thus one hole in the Fermi sea of
$\l^1_\a$. We denote the rapidity corresponding to this hole by
$\l^h$. The rapidity corresponding to the holon is denoted by
$\La^p$. The Bethe equations 
read 
\bea
\frac{2\pi}{L} I_\a &=& (2+\frac{1}{L}) \theta(\l_\a) -
\frac{1}{L}
\sum_{\b=1}^{\frac{L}{2}+1}\theta(\frac{\l_\a-\l_\b}{2})+\theta(\frac{\l_\a
+\l_\b}{2} )\nn
&& +\frac{1}{L}\kappa_{ij}^{(1)}(\l_\a)+\frac{1}{L}
\left[\theta(\frac{\l_\a-\l^h}{2})+\theta(\frac{\l_\a+\l^h}{2})+
\theta(\l_\a-\La^p)+\theta(\l_\a+\La^p)\right],\nn
\frac{2\pi}{L} J &=& \frac{1}{L}\sum_{\a=1}^{\frac{L}{2}+1}
\theta(\La^p-\l_\a)+\theta(\La^p+\l_\a)+ \frac{1}{L}\omega_{ij}(\La^p)
-\frac{1}{L}\left[\theta(\La^p-\l^h)+\theta(\La^p+\l^h)\right].
\label{baehs}
\eea
We note that the ground state at half-filling is identical to the one
of the XXX Heisenberg chain and is obtained by filling the rapidities
$\l^1_\a$ between $-\infty$ and $\infty$. 
In the limit $L\rightarrow\infty$ the distribution of roots $\l_\a$ is
described by a single integral equation for the density of roots
$\rho_s(\l)$, which is of the same structure as (3.28) of
\cite{xxz1}. The main complication is that we need to take into
account all contribution to order $\frac{1}{L}$ and thus must deal
with the fact that the roots are distributed not between $-\infty$ and
$\infty$ but between two finite,
$L$-dependent values $-A$ and $A$. The integral equation is of
Wiener-Hopf form but cannot be solved in a form sufficiently explicit
for the purpose of determining the boundary phase-shifts.
Following \cite{gmn} and \cite{saleur} we make the assumption that the
contributions due to the shift of integration boundaries will be of
higher order in $\frac{1}{L}$ as far as the boundary phase-shifts are
concerned and take $A=\infty$ (the validity of this assumption is
discussed at the end of the section). The integral equation then
can be solved by Fourier transform with the result 
\be
\widetilde{\rho_s}(\o) = 2 \gt_0(\o)+\frac{1}{L}\left\lbrace
\gt_1(\o)[1+2\cos(\o\l^h)]+\gt_0(\o)[1+2\cos(\o\La^p)]+f_{ij}(\o)
\right\rbrace ,
\label{dens} 
\ee
where $\gt_x(\o)=\frac{\exp(-\frac{x}{2}|\o|)}{2\cosh(\frac{\o}{2})}$
and where the contributions $f_{ij}$ due to the boundary fields are
\be
f_{ab}(\o)=\gt_{-1-S_1}(\o)\ ,\quad f_{ba}(\o)=\gt_{-1-S_L}(\o)\ ,\quad
f_{bb}(\o)=f_{ab}(\o)+f_{ba}(\o)\ ,\quad f_{aa}(\o)=0\ .
\ee

For the further analysis it is convenient to define counting
functions $z_s(\l)$ and $z_c(\l)$ 
\bea
z_s(\l) &=& \frac{2L+1}{2\pi} \theta(\l) - \frac{1}{2\pi}
\sum_{\b=1}^{\frac{L}{2}+1}\theta(\frac{\l-\l_\b}{2})+\theta(\frac{\l
+\l_\b}{2} )\nn
&& +\frac{1}{2\pi}\kappa^{(1)}_{ij}(\l)+\frac{1}{2\pi}
\left[\theta(\frac{\l-\l^h}{2})+\theta(\frac{\l+\l^h}{2})+
\theta(\l-\La^p)+\theta(\l+\La^p)\right],\nn
z_c(\La)&=& \frac{1}{2\pi}\sum_{\a=1}^{\frac{L}{2}+1}
\theta(\La-\l_\a)+\theta(\La+\l_\a)+ \frac{1}{2\pi}\omega_{ij}(\La)
-\frac{1}{2\pi}\left[\theta(\La-\l^h)+\theta(\La+\l^h)\right].
\label{count}
\eea
Note that for any root {\sl e.g.} $\l_\a$ of \r{baehs} the counting
function takes the integer value $z_s(\l_\a)=I_\a$ by
construction. In the thermodynamic limit $\frac{1}{L}$ times the
derivative of $z_s(\l)$ yields the distribution function of
rapidities $\rho_s(\l)$. Straightforward integration of the density
$\rho_s(\l)$ yields the following results for the counting functions
in the thermodynamic limit evaluated at the rapidities of the spinon
and holon respectively 
\bea
2\pi z_s(\l^h)&=&2Lp_s(\l^h) + \varphi_{sc}(\l^h-\La^p) + 
\varphi_{sc}(\l^h+\La^p)+\phi_s(\l^h)=0\ {\rm mod}\ 2\pi\ ,\nn
-2\pi z_c(\La^p)&=&2Lp_c(\La^p) + \varphi_{sc}(\La^p-\l^h) + 
\varphi_{sc}(\l^h+\La^p)+\phi_c(\La^p)=0\ {\rm mod}\ 2\pi\ ,
\label{cf}
\eea
where $p_{s,c}(\l)$ are the spinon/holon momenta \r{mtmhf},
$\varphi_{sc}(\l)$ is the bulk phase-shift for spinon-holon scattering
\r{smhf}, and
\bea
\phi_s(\l)&=&-\int_{-\infty}^\infty\frac{d\o}{i\o}\left[
\gt_1(\o)(1+\exp(-i\o\l))+\gt_0(\o)+f_{ij}(\o)\right]\exp(-i\o\l)\
,\nn 
\phi_c(\l)&=&\int_{-\infty}^\infty\frac{d\o}{i\o}\left[
\gt_1(\o)(1+\exp(-i\o\l))-\gt_0(\o)+f_{ij}(\o)\exp(-|\frac{\o}{2}|)\right]
\exp(-i\o\l)-\omega_{ij}(\l)\ .\nn 
\eea
The last equalities in \r{cf} hold due to the fact that evaluation of
the counting function at a root of the Bethe equations yields an
integer number.
From these equations we can now infer the boundary phase shifts by
comparing them with the quantization condition \r{qchf2}.

The scattering of spinons on a $b$-type boundary with boundary field
$h$ is identical to the one in the XXX spin chain and the results are
the same as \cite{gmn}: the phase for a spinon with spin-up and
rapidity $\l$ is given by   
\be
e^{i\psi_{s,\up}^{(b)}(\l,h)} = 
\frac{\Gamma(\frac{1}{4}-\frac{i\l}{2})}{\Gamma(\frac{1}{4}+\frac{i\l}{2})}
\frac{\Gamma(1+\frac{i\l}{2})}{\Gamma(1-\frac{i\l}{2})}
\frac{\Gamma(\frac{2-h}{4h}-\frac{i\l}{2})}
{\Gamma(\frac{2-h}{4h}+\frac{i\l}{2})}
\frac{\Gamma(\frac{2+h}{4h}+\frac{i\l}{2})}
{\Gamma(\frac{2+h}{4h}-\frac{i\l}{2})}\ .
\ee
The analogous phase for a spin-down spinon can be obtained in the
following way \cite{gmn}: As pointed out above switching to the
reference state with all spins down leads to a redefinition of the
quantities $S_j$. The excitation constructed in a way analogous to the
one above over this reference state has spin quantum number
$S^z=-\frac{1}{2}$. In order to study negative boundary fields we need
to keep track of the modification in the quantities $S_{1,L}$, which
are now always positive. Repeating the above analysis for this case we
obtain the following boundary phase-shifts for spin-down spinons
\be
e^{i\psi_{s,\da}^{(b)}(\l,h)} =
-\frac{\l+i\frac{2-h}{2h}}{\l-i\frac{2-h}{2h}}\
e^{i\psi_{s,\up}^{(b)}(\l,h)}\ .  
\ee
The phases for scattering of spinons on $a$-type boundaries are very
similar, {\sl e.g.} 
\be
e^{i\psi_{s\up}(\l,h^\prime)}=
\frac{\Gamma(\frac{1}{4}-\frac{i\l}{2})}{\Gamma(\frac{1}{4}+\frac{i\l}{2})}
\frac{\Gamma(1+\frac{i\l}{2})}{\Gamma(1-\frac{i\l}{2})}\ .
\ee
These expressions can be obtained from the $b$-type phases by setting
the boundary fields to zero. Physically this means that the spinons do
not ``see'' the $a$-type boundary fields at half-filling.

The phases $\psi_c^{(a)}$ and $\psi_c^{(b)}$ accumulated by a holon
scattering off a boundary of type $a$ or $b$ are given by
\bea
e^{i\psi_c^{(a)}(\l,h)}&=&
\left(\frac{1-i(\frac{h\l}{1-h})}{1+i(\frac{h\l}{1-h})}\right)\left(
\frac{\Gamma(\frac{3}{4}+\frac{i\l}{2})}{\Gamma(\frac{3}{4}-\frac{i\l}{2})}
\frac{\Gamma(1-\frac{i\l}{2})}{\Gamma(1+\frac{i\l}{2})}\right)
\ ,\nn
e^{i\psi^{(b)}(\l,h)}&=&
\frac{\Gamma(\frac{1+h}{2h}-\frac{i\l}{2})}
{\Gamma(\frac{1+h}{2h}+i\frac{\l}{2})}
\frac{\Gamma(\frac{1}{2h}+\frac{i\l}{2})}
{\Gamma(\frac{1}{2h}-i\frac{\l}{2})}
\left(\frac{\Gamma(\frac{3}{4}+\frac{i\l}{2})}
{\Gamma(\frac{3}{4}-\frac{i\l}{2})}\frac{\Gamma(1-\frac{i\l}{2})}
{\Gamma(1+\frac{i\l}{2})}\right)\ .
\label{hol}
\eea
Thus scattering of holons off the boundaries is influenced by both
types of boundary fields.
Let us now investigate the analytic structure of the above
phase-shifts. In \cite{faddtak} the physical strip for the spinon
rapidity was defined by the condition $|{\rm Im}(\l)|<1$ on the basis
of periodicity of the expressions for momentum and energy.
We propose the further constraint on the physical sheet that the
imaginary part of the spinon momentum has to be positive (interpreting
the spinons as {\sl particles}). This implies that the spectral
parameters should lie in the strip $0\leq {\rm Im}(\l)<1$. It now can
be seen that the spinon boundary $S$-matrices have no poles on this
strip (note that for $e^{i\psi_{s\da}}$ there is no pole at
$\l=i\frac{2-h}{2h}$). Therefore there are no boundary bound states in
the region of boundary fields we consider here. 
Let us now turn to the holon boundary $S$-matrices. 
The physical strip for the holon rapidity (for vanishing real part) is
given by $-1<{\rm Im}(\l)<0$. We see that the holon boundary
phase-shifts do not have poles on the physical sheet for the range of
boundary fields considered here ($h\in (0,2)$ and $h^\prime\in (0,1)$).
However we see that the pole $\l=i\frac{h}{1-h}$ of \r{hol} for a-type
boundaries starts crossing over onto the physical sheet for
$h^\prime\to 1$. This indicates that for $h^\prime >1$ a holon bound
state forms. Indeed we find that for $h>1$ a solution of \r{bae}
exists where one of the roots $\1l$ takes the value
$\l=i\frac{h^\prime}{1-h^\prime}$. For $h^\prime >1$ this root is
present in the ground state so that it indeed corresponds to a
boundary bound state. We see that a sufficiently strong boundary field
is needed for a bound state to be formed.

Let us now dicuss the validity of our ``incomplete'' $\frac{1}{L}$
analysis of the densities/counting functions. As pointed out above a
complete analytical treatment encounters significant technical
difficulties. However our results can be checked numerically in the
following way: if the shift in integration boundary would lead to an
additional term of order $\frac{1}{L}$, \r{cf} should be incorrect to
order one. We performed an explicit numerical evaluation of $\La^p$
and $\lambda^h$ corresponding to integers $L/n$ and $L/m$ with $n$ and
$m$ fixed by solving the Bethe equations \r{baehs} for chains up to
$600$ sites. Through insertion of the numerical values of $\l^h$ and
$\La^p$ into \r{cf} it is then possible to check whether the neglected
finite-size effects contribute to order one in the counting
functions. We found no evidence for any missing contribution to the
boundary phase-shifts. We therefore conclude that for the half-filled
band the shift in integration boundary can indeed be neglected.

\subsection{\sc Boundary S-matrix for the less than Half-Filled Band}

Let us now turn to the case of the less than half-filled band. 
It is a well-known fact that below half-filling the scattering
matrices are not functions of the differences of the rapidities of the
scattering particles any longer. Therefore we need to replace \r{qc}
by an appropriately generalized prescription. This can be accomplished
by following through the arguments used in \cite{saleur} to derive
\r{qc}: first of all the scattering considered below is diagonal in
the sense that whenever a spinon or holon bounces off a wall, it
merely changes its rapidity $\l$ to $-\l$ and acquires a
phase-shift. Secondly the wave-functions of the excitations on our
finite interval with fixed boundary conditions are standing waves of
states with opposite rapidities, which leads to the condition
\be
\exp(iLp(\l_1))S(\l_1,\l_2)K_1(\l_1,h_1)=
\exp(-iLp(\l_1))S(-\l_1,\l_2)K_1(-\l_1,h_L)\ ,
\label{qc2}
\ee
where $p(\l)$ is again the physical momentum of the corresponding
(infinite) periodic system and $S(\l,\nu)$ are the bulk scattering 
matrices. The extraction of the boundary phase-shifts from evaluating
the counting functions at the spectral parameters of the
holons/spinons is much more problematic than for the half-filled case as
now the integration boundary $B$ corresponding to the charge Fermi sea
is finite and the issue of how to treat the $\frac{1}{L}$ corrections
to $B$ arises.

\section{\sc Conclusion}

In this work we have studied zero-temperature boundary effects in an
open supersymmetric $t$-$J$ chain with boundary fields. Surface 
contributions to ground state properties were evaluated as functions
of the boundary fields and the phase-shifts acquired by holons and
spinons scattering off a boundary were determined. It also would be
interesting to extend the analysis to finite temperatures. This
appears to be difficult as among other things the usual expression for
the entropy \cite{yayatba} has to be modified in order to project out
the spurious states. The best way to deal with these problems seems to
be a Thermal Bethe Ansatz analysis \cite{TBA}. \\  
\indent 
The (subleading) finite-size corrections to the ground state energy,
and energies of low-lying excited states, can be evaluated by means of
the Euler-Maclaurin sum formula as well. This was done for the
case of the $aa$-boundary conditions and zero bulk magnetic field in
\cite{suz}. That analysis, which can be straightforwardly extended to
the other sets of integrable boundary conditions considered here,
allows to make contact with Conformal Field Theory in a geometry with
boundary \cite{cardy}.

\vspace*{1cm}

\begin{center}
{\large \sc Acknowledgements}
\end{center}
I am grateful to A. Tsvelik, R. Nepomechie, L. Mezincescu, H. Frahm
and J.-S. Caux for helpful discussions and suggestions.
This work was supported by the EU under Human Capital and
Mobility fellowship grant ERBCHBGCT940709.

\appendix
\section{}
In this appendix we derive the expression \r{E} for the ground state
energy. The ground state for given magnetic field $H$ and chemical
potential $\mu$ is obtained by filling all vacancies for the integers
$I^1_\a$ from $1$ to $I_{\rm max}=N_\da+N_h$  and all vacancies for
the integers $J_\g$ between $1$ and $J_{\rm max}=N_h$. Inserting this
description into the Bethe equations \r{baelog} and then subtracting
susequent equations for $\a$ and $\a +1$ and $\g$ and $\g +1$ we
obtain the following equations for the densities
$\vr_s(\l_\a)=\frac{1}{L(\l_{\a +1}-\l_\a)}$ and
$\vr_c(\1l_\g)=\frac{1}{L(\1l_{\g +1}-\1l_\g)}$ 
\bea
\vr_s(\l_\a)&=& 2a_1(\l_\a) -
\frac{1}{L} \sum_{\b}a_2(\l_\a-\l_\b)+a_2(\l_\a+\l_\b)
+\frac{1}{L} \sum_{\gamma}a_1(\l_\a-\1l_\g)+a_1(\l_\a+\1l_\g)\nn
&&+\frac{1}{L}\left(\frac{\kappa^\prime_{ij}(\l_\a)}{2\pi}
+a_1(\l_\a)\right)\nn
\vr_c(\1l_\g) &=&
\frac{1}{L}\sum_{\alpha}a_1(\1l_\g-\l_\a)+a_1(\1l_\g+\l_\a)+
\frac{1}{2\pi L}\omega^\prime_{ij}(\1l_\g)\ .
\label{Appdens}
\eea
Here $\kappa^\prime_{ij}(\l_)$ and $\omega^\prime_{ij}(\l)$ are
the derivatives of $\kappa_{ij}^{(1)}(\l)$ and $\omega_{ij}(\l)$
defined in \r{bc} respectively. Now we follow \cite{saleur}
and rewrite \r{Appdens} in terms of a set of ``doubled'' variables
\bea
\nu_\a&=&\cases{-\l_{N_\da+H_h-\a} & $\a=0,\ldots, N_\da+N_h$\cr
\l_{\a-N_\da-N_h}  & $\a=N_\da+N_h+1,\ldots, 2(N_\da+N_h)$\cr}\nn
\nu^{(1)}_\g &=&\cases{-\l^{(1)}_{N_h-\g} & $\g=0,\ldots N_h$\cr
\l_{\g-N_h}  & $\g=N_h+1,\dots 2N_h\ ,$\cr}
\eea
where we defined $\l_0=0$ and $\l^{(1)}_0=0$.
Now we take the thermodynamic limit of the equations \r{Appdens}
written in the new variables. This is done by using the
Euler-Maclaurin sum formula to turn sums into integrals (see {\sl e.g}
\cite{eml,xxz1,hol}). Care has to be exercised in order to take into
account the fact that terms depending on the spectral parameters
located at zero must be subtracted explicitly. After some
manipulations we arrive at following coupled integral equations for
the densities $\rho_s(\nu_\a)=\frac{1}{L(\nu_{\a +1}-\nu_\a)}$ and
$\rho_c(\nu^{(1)}_\g)=\frac{1}{L(\nu^{(1)}_{\g +1}-\nu^{(1)}_\g)}$ 
\bea
\rho_s(\l)&=& 2a_1(\l) -
\int_{-A^+}^{A^+}\!\!d\mu\ a_2(\l-\mu)\ \rho_s(\mu)
+\int_{-B^+}^{B^+}\!\!d\mu\ a_1(\l-\mu)\ \rho_c(\mu)
+\frac{1}{L}\left(\frac{\kappa^\prime_{ij}(\l)}{2\pi}
+a_2(\l)\right)\nn
\rho_c(\l) &=&
\int_{-A^+}^{A^+}\!\!d\mu\ a_1(\l-\mu)\ \rho_s(\mu) 
+\frac{1}{L}\left(\frac{\omega^\prime_{ij}(\l)}{2\pi}-a_1(\l)\right)\
,
\label{Adens}
\eea
where $A^+$ and $B^+$ are the spectral parameters corresponding to the
maximal taken integers $I^1_\a$ and $J_\g$ plus $\frac{1}{2}$. Higher
order terms in the Euler-Maclaurin expansion have been dropped as they
turn out to not contribute to the surface energy. 

The energy per site \r{bareE} in the thermodynamic limit can be
expressed in terms of the densities $\rho_{s,c}$ as\footnote{We
again turn the sums into integrals by means of the Euler-Maclaurin
formula.} 
\bea
\frac{E}{L}&=&-\int_{-{A^+}}^{A^+}d\nu\ [\pi a_1(\nu)-\frac{H}{2}]\ 
\rho_s(\nu)+\int_{-{B^+}}^{B^+}d\nu\ [\frac{\mu}{2}-\frac{H}{4}]\
\rho_c(\nu)\nn 
&&+\frac{1}{L}\left(-\frac{\mu}{2}-\frac{H}{4}+E_{ij}+2\right)
-\mu-\frac{H}{2}.
\label{Appen}
\eea
Note that we have divided the bare energies by two due to the fact
that we are working with the densities of the doubled variables. We
also explicitly subtracted a term
$\frac{1}{L}(\frac{\mu}{2}+\frac{H}{4}-2)$ to compensate for the
spurious roots at zero. To proceed further we rewrite \r{Adens} as an
operator equation 
\be
\left(\matrix{I-\widehat{K}_{ss}&-\widehat{K}_{sc}\cr
-\widehat{K}_{cs}& I-\widehat{K}_{cc}\cr}\right)*\pmatrix{\rho_s
\cr\rho_c\cr}=\pmatrix{\rho^{(0)}_s\cr\rho^{(0)}_c\cr}\ ,
\label{opeq}
\ee
where $I$ is the identity operator and $*$ denotes convolution with
the appropriate kernel {\sl e.g.} 
$-\widehat{K}_{ss}*\rho_s\bigg|_\l = \int_{-{A^+}}^{A^+}d\mu\ a_2(\l-\mu)\
\rho_s(\mu)$ and where
\be
\rho^{(0)}_s =
2a_1(\l)+\frac{1}{L}\left(\frac{\kappa^\prime_{ij}(\l)}{2\pi} 
+a_2(\l)\right)\ ,\quad 
\rho^{(0)}_c = \frac{1}{L}\left(\frac{\omega^\prime_{ij}(\l)}{2\pi}
-a_1(\l)\right)\ .
\ee

We note that the above integral kernels are all symmetric. In what
follows we use the shorthand notation $(id-\widehat{K})_{ab}\rho_b=
\rho^{(0)}_a$ for \r{opeq}. Let us now define quantities $\eps_s(\l)$
and $\eps_c(\l)$ through the integral equations
$(id-\widehat{K})_{ab}\eps_b=\eps^{(0)}_a$, where $\eps^{(0)}_s(\l) =
-2\pi a_1(\l) +H$ and $\eps^{(0)}_c(\l) = \mu -\frac{H}{2}$.
With $\chi = -\frac{\mu}{2}-\frac{H}{4}+E_{ij}+2$ the energy per site
can now be written as 
\bea
\frac{E}{L}&=& \frac{1}{2}\sum_{b=s,c}\int_{-C^+_b}^{C^+_b}d\mu\
\eps_b^{(0)}(\mu)\rho_b(\mu)-\mu-\frac{H}{2}+\frac{\chi}{L}\nn
&=&\frac{1}{2}\sum_{a,b=s,c}\int_{-C^+_b}^{C^+_b}d\mu\
\eps_b^{(0)}(\mu) (id-\widehat{K})^{-1}_{ba}*\rho^{(0)}_a\bigg|_\mu
-\mu-\frac{H}{2}+\frac{\chi}{L}\nn
&=&\frac{1}{2}\sum_{a=s,c}\int_{-C^+_a}^{C^+_a}d\mu\
\eps_a(\mu) \rho^{(0)}_a(\mu)-\mu-\frac{H}{2}+\frac{\chi}{L}\ ,
\label{appe}
\eea
where $C^+_s=A^+$ and $C^+_c=B^+$. In the thermodynamic limit the
ground state is obtained by minimizing $E$ with respect to the
integration boundaries $A^+$ and $B^+$ (see {\sl e.g.} \cite{hol}),
{\sl i.e.} $\frac{\partial E}{\partial C^+_a}\bigg|_{C^+_a=C_a} = 0$,
where $C_s=A$, $C_c=B$. From this fact it follows that the integration
boundaries $C^+_a$ in \r{appe} can be replaced by $C_a$ with error of
${\cal O}(L^{-2})$, which does not affect the surface energy we are
after. In other words we can replace $C_a^+$ by $C_a$ (up to ${\cal
O}(L^{-2})$) due to the fact that the dressed energies vanish at the
integration boundaries. This finally establishes \r{E}.

\section{}
In this appendix we outline how to evaluate the integral \r{epsb}. We
first note that {\sl via} Fourier transform the following equality can
be established
\bea
\int_{-A}^Ad\mu\ a_S(\mu)\ \eps_s(\mu) &=& -2\pi G_S(0)+\frac{H}{2} 
+ 2\int_0^\infty d\mu\left[G_{1+S}(\mu+A)-a_S(\mu+A)\right]y(\mu)\nn
&&+ \int_{-B}^B G_S(\mu)\ \eps_c(\mu)\ .
\eea
The last term on the r.h.s. is readily evaluated by using the
fact that $G_S(\mu)$ is smooth around zero and thus can be Taylor
expanded 
\be
\int_{-B}^B G_S(\mu)\ \eps_c(\mu) = -2\zeta(3) B^3 G_S(0)\ .
\ee
The second term on the r.h.s. (which will be denoted by $R_2$ in the
following) is more difficult to treat. In the
region where $S\ll A$ we can use the asymptotic expansions
$G_{1+S}(\mu+A)\sim \frac{1+S}{4\pi(\mu+A)^2}$ and
$a_{S}(\mu+A)\sim \frac{S}{2\pi(\mu+A)^2}$ to determine the leading
contribution to the integral. We then Laplace transform
$\frac{1}{(\mu+A)^2}$ and after some manipulations arrive at 
\be
\int_0^\infty d\mu\ \frac{1}{(A+\mu)^2}\ y(\mu) = \int_0^\infty dx\
xe^{-Ax} \widetilde{y}^+(ix)\ .
\ee
Due to the strongly decaying factor $e^{-Ax}$ the leading contribution
to this integral clearly comes from the region around $x=0$. Expanding
$\widetilde{y}^+(ix)$ around $x=0$ we finally obtain
\be
R_2=\frac{S-1}{2}\frac{H}{\ln H}+\ {\rm subleading\ terms}\ .
\ee
In the region $S\gg A$ the above strategy for determining $R_2$ is not
applicable. Instead we Fourier transform and arrive at
\be
R_2 = -\frac{1}{2\pi A}\int_0^\infty d\o\ 
\frac{e^{-\frac{S-1}{2A}\o}}{\cosh(\frac{\o}{2A})}\left\lbrace
\cos\o\left(\widetilde{y}^+(\frac{\o}{A})
+\widetilde{y}^+(-\frac{\o}{A})\right)
+i\sin\o\left(\widetilde{y}^+(\frac{\o}{A})
-\widetilde{y}^+(-\frac{\o}{A})\right)\right\rbrace .
\ee
We again have a strongly decaying factor in the integrand (recall that
$\frac{S-1}{2A}\gg 1$) so that we can expand the other terms around
$\o=0$. The problem we run into now is that the subleading terms in
the expansion of $y(\l)$ contribute in an important way. The
leading contribution of $y_1$ to $R_2$ is found to be (after expanding
$\widetilde{y}^+_1$ around $\o=0$ only elementary integrals are
encountered) 
\be
-\frac{H}{2}+\frac{H}{\pi^2}\frac{\ln(S-1)}{S-1}-\frac{2H\ln
H}{\pi^2(S-1)}\ .
\ee
The second term is precisely cancelled by the leading contribtuion
from $y_2$ to $R_2$, whereas the further terms are all small compared
to $-\frac{2H\ln H}{\pi^2(S-1)}$.


\begin{thebibliography}{10}
\begin{center}
{\large\sc References}
\end{center}
\baselineskip=12pt
\bibitem{tsv}
P. de Sa, A. Tsvelik,
\newblock\PRB{52}{1995}{3067}.

\bibitem{tsv0}
A. Tsvelik,
\newblock\JPA{28}{1995}{L625}.

\bibitem{bares0}
P.A. Bares, {\sl preprint} cond-mat/9412011.

\bibitem{saleur}
P. Fendley, H. Saleur,
\newblock \NPB{428}{1994}{681}.

\bibitem{henrik}
P. Fr\"ojdh, H. Johannesson, \PRL{75}{1995}{300}.

\bibitem{fls}
P. Fendley, A.W.W. Ludwig, H. Saleur, {\sl preprint} cond-mat/9503172. 

\bibitem{kf}
C.L. Kane, M.P.A. Fisher, \PRB{46}{1992}{7268}.

\bibitem{aff}
I. Affleck, {\sl preprint} cond-mat/9512099. 

\bibitem{che}
I.V. Cherednik,
\newblock \TMP{61}{1984}{977}.

\bibitem{xxz2}
F.C. Alcaraz, M.N. Barber, M.T. Batchelor, R.J. Baxter,
G.R.W. Quispel, \JPA{20}{1987}{6397}.

\bibitem{skl}
E.K. Sklyanin,
\newblock \JPA{21}{1988}{2375}.

\bibitem{mn1}
L. Mezincescu, R.I. Nepomechie, 
\newblock \JPA{24}{1991}{L17}, \JPA{25}{1992}{2533}.

\bibitem{xxz1}
C.J. Hamer, G.R.W. Quispel, M.T. Batchelor, \JPA{20}{1987}{5677}.

\bibitem{schulz}
H. Schulz, \JPC{18}{1985}{581}.

\bibitem{gr}
A. Gonzales-Ruiz,
\newblock \NPB{424}{1994}{468}.

\bibitem{fk2}
A. F\"orster, M. Karowski,
\newblock \NPB{408}{1993}{512}.

\bibitem{fk}
A. F\"orster, M. Karowski,
\newblock \NPB{396}{1993}{611}.

\bibitem{suth}
B. Sutherland,
\newblock\PRB{12}{1975}{3795}.

\bibitem{schlott}
P. Schlottmann,
\newblock \PRB{36}{1987}{5177}.

\bibitem{bcfh}
P.A. Bares, J.M.P. Carmelo, J. Ferrer, P. Horsch,
\newblock \PRB{46}{1992}{14624}.

\bibitem{bares1}
P.A. Bares,
\newblock PhD thesis, ETH Z\"urich (1991)

\bibitem{bares2}
P.A. Bares, G.Blatter, M.Ogata,
\newblock \PRB{44}{1991}{130}.

\bibitem{sarkar}
S. Sarkar,
\newblock\JPA{24}{1991}{1137}.

\bibitem{ektj}
F.H.L. E\ss ler, V.E. Korepin,
\newblock \PRB{46}{1992}{9147}.

\bibitem{vladb}
V.E. Korepin, N.M. Bogoliubov, A.G. Izergin,
\newblock {\sl Quantum Inverse Scattering Method, Correlation
Functions and Algebraic Bethe Ansatz}, Cambridge University Press, 1993

\bibitem{ek}
F.H.L. E\ss ler, V.E. Korepin,
\newblock \IJMPB{8}{1994}{3243}

\bibitem{eks}
F.H.L. E\ss ler, V.E. Korepin, K. Schoutens,
\newblock \IJMPB{8}{1994}{3205}

\bibitem{holger}
H. Frahm, V.E. Korepin,
\newblock \PRB{43}{1991}{5653}.

\bibitem{nep1}
L. Mezincescu, R.I. Nepomechie, 
\newblock {\sl cond-mat 9208021}.

\bibitem{yaya}
C.N. Yang, C.P. Yang,
\newblock \PR{150}{1966}{327}.

\bibitem{kondo}
A. Tsvelik, P. Wiegmann, \AdP{32}{1983}{453},\\
N. Andrei, K. Furuya, J.H. Lowenstein, \RMP{55}{1983}{331}.

%\bibitem{ekh1}
%F.H.L. E\ss ler, V.E. Korepin,
%\newblock \PRL{72}{1994}{908}

\bibitem{ekh2}
F.H.L. E\ss ler, V.E. Korepin,
\newblock \NPB{426}{1994}{505}

\bibitem{barry}
B.M. McCoy, {\sl private communication}, see also p. 3618 of
R. Kedem, B.M. McCoy, \IJMPB{8}{1994}{3601}.

\bibitem{korepin}
V.E. Korepin,
\newblock \TMP{41}{1979}{169}

\bibitem{gmn}
M.T. Grisaru, L. Mezincescu, R.I. Nepomechie, 
\newblock \JPA{28}{1995}{1027}.

\bibitem{al}
N. Andrei, J.H. Lowenstein,
\newblock \PLA{80}{1980}{401}.

\bibitem{ad}
N. Andrei, C. Destri,
\newblock \NPB{231}{1984}{455}.

\bibitem{faddtak}
L.D. Faddeev, L. Takhtajan,
\newblock \JSM{24}{1984}{241}.

\bibitem{yayatba}
C.N. Yang, C.P. Yang,
\newblock \JMP{10}{1969}{1115}.

\bibitem{TBA}
T. Koma, \PTP{78}{1987}{1213},\\
M. Takahashi, \PRB{43}{1991}{5788},\\
C. Destri, H. deVega, \NPB{438}{1995}{413}.

\bibitem{suz}
H. Asakawa, M. Suzuki,
{\sl ``Finite size corrections in the supersymmetric $t$-$J$ model
with boundary fields''}, preprint.

\bibitem{cardy}
J. Cardy,
\NPB{240}{1984}{514}.

\bibitem{eml}
F. Woynarovich, H.-P. Eckle,
\JPA{20}{1987}{L97}.

\bibitem{hol}
F. Woynarovich, \JPA{22}{1989}{4243},\\
H. Frahm, N.-C. Yu, \JPA{23}{1990}{2115}.
\end{thebibliography}
\end{document}